\documentclass[onecolumn,a4paper,CEJP,PDF]{elsart}
\usepackage{epsfig,graphicx,amssymb,amsmath,cite,makeidx}
\usepackage{layout}
\usepackage{amsmath}
\usepackage{textcomp}
\usepackage{hyperref}
\usepackage{latexsym}
\usepackage{graphicx}
\usepackage{epsfig}
\usepackage{float}
\journal{CEJP}
\begin{document}
\begin{frontmatter}
\title{Improved Eavesdropping Detection Strategy in Quantum Direct Communication Protocol Based on Four-particle GHZ State}
\author[1]{Li Jian},
\author[1]{Jin Haifei\corauthref{cor}},
\corauth[cor]{E-mail:jinhaifei@bupt.edu.cn}
\author[1,2]{Jing Bo}
\begin{flushleft}{\footnotesize
\emph{$^1$ School of Computer, Beijing University of Posts and Telecommunications,\\
 Beijing 100876, People's Republic of China\\
$^2$ Department of Computer Science, Beijing Institute of Applied Meteorology,\\
 Beijing 100029, People's Republic of China\\}}
\end{flushleft}
\begin{abstract}In order to improve the eavesdropping detection efficiency in two-step quantum direct communication protocol, an improved eavesdropping detection strategy using four-particle GHZ state is proposed, in which four-particle GHZ state is used to detect eavesdroppers. During the security analysis, the method of the entropy theory is introduced, and two detection strategies are compared quantitatively by using the constraint between the information which eavesdropper can obtain and the interference introduced. If the eavesdroppers intend to obtain all information, the eavesdropping detection rate of the original two-step quantum direct communication protocol by using EPR pair block as detection particles is 50\%; while the proposed strategy's detection rate is 88\%. In the end, the security of the proposed protocol is discussed. The analysis results show that the eavesdropping detection strategy presented is more secure.
\end{abstract}
\end{frontmatter}
\section{Introduction}
\noindent The goal of cryptography is to ensure that the secret message is intelligible only for the two authorized parties of communication and should not be altered during the transmission. So far, it is trusted that the only proven secure cryptophyte is the one-time-pad scheme in which the secret key is as long as the message. The two distant parties who want to transmit their secret message must distribute the secret key first. But it is difficult to distribute securely the secret key through a classical channel. The quantum key distribution (QKD), whose task is to create a secret key between two remote authorized users, is one of the most remarkable applications of quantum mechanics and the only proven protocol for secure key distribution. Since Bennet and Brassard presented the pioneer QKD protocol (BB84 protocol) [1] in 1984, a lot of quantum information security processing methods have been advanced, such as quantum teleportation [2-7], quantum dense coding [8-9], quantum secret sharing [10-11] and so on.\\
\indent In recent years, a novel concept, quantum secure direct communication (QSDC) was put forward and studied by some groups. Different from key distribution whose object is to establish a common random key between two parties, a secure direct communication is to communication important message directly without first establishing a random key to encrypt them. Thus secure direct communication is more demanding on the security. As a secure direct communication, it must satisfy two requirements. First, the secure message should be read out directly by the legitimate user Bob when he receives the quantum state and no additional classical information is needed after the transmission of particles. Second, the secret message which has been encoded already in the quantum states should not leak even though an eavesdropper may get hold of the channel. That is to say, the eavesdropper cannot only be detected but also obtains blind results. As classical message can be copied fully, it is impossible to transmit secret message directly through classical channels. But when quantum mechanics enters into the communication, the story will change.\\
\indent Another class of quantum communication protocols [12-14] used to transmit secret message are called deterministic secure quantum communication (DSQC). Certainly, the receiver can read out the secret message only after he ex-changes at least one bit of classical information for each particle with the sender in a DSQC protocol, which is different from QSDC. DSQC is similar to QKD, but it can be used to obtain deterministic information, not a random binary string, which is different from the QKD protocols in which the user cannot predict whether an instance is useful or not.\\
\indent Many people are interested in researching QSDC, and many protocols like QSDC were proposed, including the protocols without using entanglement [15-17], the protocols using entanglement [18-23] and the two-way QSDC protocols [24-33].The QSDC protocol can also be used in some special environments as first proposed by Bostr?m et al. [34] and Deng et al. [18]. In Ref. [34], Bostrom and Felbinger presented a famous QSDC protocol which is called "ping-pong" protocol. But researchers have found much vulnerability in the "ping-pong" protocol, such as the "ping-pong" protocol cannot resist the "man-in-middle attack" and the transmission capacity is low.\\
\indent In order to improve the eavesdropping detection efficiency in two-step quantum direct communication protocol, an improved eavesdropping detection strategy using four-particle GHZ state is proposed, in which four-particle GHZ state is used to detect eavesdroppers. During the security analysis, the method of the entropy theory is introduced, and two detection strategies are compared quantitatively by using the constraint between the information which eavesdropper can obtain and the interference introduced. If the eavesdroppers intend to obtain all information, the eavesdropping detection rate of the original two-step quantum direct communication protocol by using EPR pair block as detection particles is 50\%; while the proposed strategy's detection rate is 88\%. In the end, the security of the proposed protocol is discussed. The analysis results show that the eavesdropping detection strategy presented is more secure.\\
\indent For simplicity, suppose that the protocol presented in Ref. [18] is shortened as DPP and the improved protocol in this paper is shortened as FPP.
\section{DPP Protocol}
\noindent An EPR pair can be in one of the four Bell states,
\begin{equation}\label{1}
\big|\psi^-\big>=\frac{1}{\sqrt{2}}(\big|01\big>-\big|10\big>),
\end{equation}
\begin{equation}\label{2}
\big|\psi^+\big>=\frac{1}{\sqrt{2}}(\big|01\big>+\big|10\big>),
\end{equation}
\begin{equation}\label{3}
\big|\phi^-\big>=\frac{1}{\sqrt{2}}(\big|00\big>-\big|11\big>),
\end{equation}
\begin{equation}\label{4}
\big|\phi^+\big>=\frac{1}{\sqrt{2}}(\big|00\big>+\big|11\big>).
\end{equation}
If the state of a single photon be measured, the Bell state will collapse and the state of the other particle will be completely determined if we know the measurement result of the first photon. As is known to all, the basic principle of the original "ping-pong" protocol is that one bit information can be encoded in the states $\big|\psi^\pm\big>$, which is completely unavailable to anyone who has access to either of the particles. To extract secret message from Alice, Bob must own both particles, for no experiment performed on only one particle can distinguish these states from each other [34].\\
\indent Let us start with a brief description of the DPP protocol.\\
(S1) Alice prepares an ordered \emph{N} EPR pairs in state $\big|\psi^-\big>$ , extracts all the first particles, and forming the sequence $S_1$ in order. The remainder particles are formed the sequence $S_2$ in order.\\
(S2) Alice sends the sequence $S_1$ to Bob. Alice and Bob then check eavesdropping by the following procedure:(a) Bob chooses randomly a number of the photons from the sequence $S_1$ and tells Alice which particles he has chosen. (b) Bob chooses randomly one of the two sets of MBs, say, $\sigma_Z$ and $\sigma_X$ to measure the chosen photons. (c) Bob tells Alice which MB he has chosen for each photon and the outcomes of his measurements. (d) Alice uses the same MB as Bob to measure the corresponding photons in the sequence $S_2$ and checks with the results of Bob. If no eavesdropper exists, their results should be completely opposite. This is the first eavesdropping check. After that, if the error rate is small, Alice and Bob can conclude that there is no eavesdropper in the line. Alice and Bob continue to perform step(S3); otherwise, they have to discard their transmission and abort the communication.\\
(S3) Alice encodes her messages on the sequence $S_2$ and transmits it to Bob. Before the transmission, Alice must encode the EPR pairs. In order to guard for eavesdropping in this transmission, Alice has to add a small trick in the sequence $S_2$. She selects randomly in the sequence $S_2$ some particles and performs on them randomly one of the four operations. The number of such particles is not big as long as it can provide an analysis of the error rate. Only Alice knows the positions of these sampling particles and keeps them secret until the communication is completed. The remaining sequence $S_2$ particles are used to carry the secret message directly. To encode the message, they use the dense coding scheme of Bennett and Wiesner [8], where the information is encoded on an EPR pair with a local operation on a single qubit. Here, the dense coding idea was generalized into secure direct communication. Different from dense coding, in this protocol, both the particles in an EPR pair are sent from Alice to Bob in two steps, and the transmission of EPR pairs is done in block. Explicitly, Alice makes one of the four unitary operations ($U_0$,$U_1$,$U_2$ and $U_3$) to each of her particles,
\begin{equation}\label{5}
U_0=I=\big|0\big>\big<0\big|+\big|1\big>\big<1\big|,
\end{equation}
\begin{equation}\label{6}
U_1=\sigma_z=\big|0\big>\big<0\big|-\big|1\big>\big<1\big|,
\end{equation}
\begin{equation}\label{7}
U_2=\sigma_x=\big|1\big>\big<0\big|+\big|0\big>\big<1\big|,
\end{equation}
\begin{equation}\label{8}
U_3=-i\sigma_y=\big|1\big>\big<0\big|-\big|0\big>\big<1\big|.
\end{equation}
And they transform the state $\big|\psi^-\big>$ into $\big|\psi^-\big>$, $\big|\psi^+\big>$ , $\big|\phi^-\big>$ and $\big|\phi^+\big>$, respectively. These operations correspond to 00, 01, 10 and 11, respectively.\\
(S4) After the transmission of sequence $S_2$, Alice tells Bob the positions of the sampling pairs and the type of the unitary operations on them. Bob performs Bell-basis measurement on the sequence $S_1$ and $S_2$ simultaneously. By checking the sampling pairs that Alice has chosen, he will get an estimate of the error rate in the sequence $S_2$ transmission. In fact, in the second transmission, Eve can only disturb the transmission and cannot steal the information because she can only get one particle from an EPR pair.\\
(S5) If the error rate of the sampling pairs is reasonably low, Alice and Bob can then entrust the process, and continue to correct the error in the secret message using error correction methods. Otherwise, Alice and Bob abandon the transmission and repeat the procedure from the beginning.\\
(S6) Alice and Bob do error correction on their results. This procedure is exactly the same as that in QKD. However, to preserve the integrity of the message, the bits preserving correction code, such as CASCADE [35], should be used.
\section{FPP Protocol}
\subsection{The process of the FPP protocol}
\noindent In the protocol presented in Ref.[36], the transmission is managed in batches of \emph{N} EPR pairs. An advantage of block transmission scheme is that we can check the security of the transmission by measuring some of the decoy photons [37-38] in the first step, where both Alice and Bob contain a particle sequence at hand, which means that an eavesdropper has no access to the first particle sequence, then no information will be leaked to her whatever she has done to the second particle sequence. Following this method using block transmission, the FPP scheme is proposed, shown in Fig.1.\\
\begin{figure}[H]
  \centering
  \includegraphics[width=5.5cm]{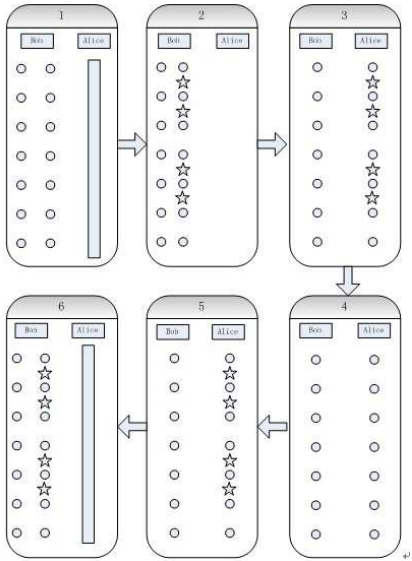}\\
  \caption{The process of the FPP}\label{FIG1}
\end{figure}
Define
\begin{equation}\label{9}
\big|\psi\big>=\frac{1}{\sqrt{2}}(\big|0000\big>+\big|1111\big>).
\end{equation}
Now let us give an explicit process for the FPP.\\
(S1) Bob prepares a large enough number (\emph{N}) of Bell states $\big|\phi^+\big>$ in order. He extracts all the first particles in these Bell state, forming the sequence \emph{A} (\emph{the travel qubits}) in order, used to transmit secure message, and the remaining particles forming the sequence \emph{B} (\emph{the home qubits}) in order.\\
(S2) Bob prepares a large number (\emph{cN/(1-c)}) of four-particle GHZ states $\big|\psi\big>$ and forms the sequence \emph{C} to detect eavesdropping. Here, \emph{c} expresses the probability of switching to the control mode in the original "ping-pong" protocol [34].Note that the sequence \emph{C} includes \emph{4cN/(1-c)} particles. In the sequence \emph{C}, Bob reserves the particle 1 of the four-particle GHZ state, and measures them by Z-basis $B_Z=\{\big|0\big>,\big|1\big>\}$.After that, Bob inserts particles 2, 3, 4 of the four-particle GHZ state to the sequence \emph{A} randomly, forming a new sequence \emph{D}, which includes decoy photons of four-particle GHZ state, but only Bob knows the position of decoy photons.\\
(S3) Bob stores the sequence \emph{B} and sends the sequence \emph{D} to Alice.\\
(S4) After Alice received the sequence \emph{D}, Bob tells her the positions where are the decoy photons and the measurement of particles 1 of four-particle GHZ state in \emph{C}. Then, Alice extracts the decoy photons from the sequence \emph{D} and performs measurement. This is the first eavesdropping check.If there is no eavesdropper, when Bob's measurement is $\big|0\big>$,then the measurement result of particles 2, 3, 4 should be $\big|000\big>$;while Bob's measurement is $\big|1\big>$, particles 2, 3, 4 should be$\big|111\big>$ , and they continue to the next step(S5), the FPP protocol keeping on. Otherwise, the communication is interrupted, and the FPP protocol switches to (S1).\\
(S5) Alice discards the decoy photons, then the sequence \emph{D} becomes to the sequence \emph{A} again. Alice encodes her messages on the sequence \emph{A} and transmits it to Bob. In order to guard for eavesdropping in this transmission, Alice also has to insert some four-particle GHZ state particles in the sequence \emph{A} before the transmission. Alice only inserts particles 2, 3, 4 of the four-particle GHZ state in the sequence \emph{A} and reserves the particle 1. Only Alice knows the positions of these decoy photons and the measurement results of the particles 1, and keeps them until the communication is completed. The sequence \emph{A} are used to carry the secret message directly. To increase the transmission capacity, the dense coding scheme be used to encode the secret message. Different from dense coding, in this protocol, the transmission of EPR pairs is done in block. Explicitly, Alice makes one of the four unitary operations ($U_0$,$U_1$,$U_2$ and $U_3$) to each of her particles, and they transform the state $\big|\phi^+\big>$ into $\big|\phi^+\big>$, $\big|\phi^-\big>$ , $\big|\psi^+\big>$ and $\big|\psi^-\big>$, respectively. These operations correspond to 00, 01, 10 and 11, respectively. Then Alice transmits the sequence \emph{A} which carries decoy photons to Bob.\\
(S6) After transmitting the sequence \emph{A}, Alice tells Bob the positions of the decoy photons and the measurement results of the particles 1. To obtain the secret message, Bob performs Bell-basis measurement on the sequences \emph{A} and \emph{B} simultaneously. By checking the decoy photons that Alice insert, Bob will get an estimate of the error rate in the sequence \emph{A} transmission. In fact, Eve can only disturb the transmission and cannot steal the information because she can only get one particle from an EPR pair. If the error rate of the decoy photons is reasonably low, Alice and Bob can then entrust the process, and continue to transmit the secret message. Otherwise, Alice and Bob abandon the transmission and repeat the procedures from the beginning.\\
\indent As discussed above, the secret message can be transmitted securely between Alice and Bob, and the eavesdropper will be found out if she disturbs the quantum line. Eve cannot read out the information from the EPR pairs even if she captures the sequence \emph{A}, because no one can read the information from one particle of the EPR pair alone. So, the improved protocol is secure.\\
\subsection{The security analysis of the protocol}
\noindent In the original "ping-pong" protocol, the author calculated the maximal amount of the information $I(d_{lO})$ that Eve can eavesdrop and the probability $d_{lO}$ that Eve is detected [34]. And the function $I(d_{lO})$ is provided. When $p_0=p_1=0.5$,
\begin{equation}\label{10}
I(d_{lO})=-d_{lO}log_2d_{lO}-(1-d_{lO})log_2(1-d_{lO}).
\end{equation}
The above method can be used to compare the efficiency of eavesdropping detection between the two protocols.\\
\indent Now, let us analyze the efficiency of eavesdropping detection in FPP protocol. In order to gain the information that Alice operates on \emph{the travel qubits}, Eve performs the unitary attack operation \emph{E} on the composed system firstly. Then Alice takes a coding operation on \emph{the travel qubits}. Eve performs a measurement on the composed system at last. Note that, all transmitted particles are sent as block before detecting eavesdropping, which is different from the original "ping-pong" protocol. For Eve does not know which particles are used to detect eavesdropping, so what she can do is only performing the same attack operation on all the particles. As for Eve, the state of \emph{the travel qubits} is indistinguishable from the complete mixture, so all \emph{the travel qubits} are considered in either of the states $\big|0\big>$ or $\big|1\big>$ with equal probability $p=0.5$.\\
\indent Generally speaking, suppose there is a group of decoy photons at the four-particle GHZ state $\big|\psi\big>$, and after performed the attack operation \emph{E}, the states $\big|0\big>$ and $\big|1\big>$ become
\begin{equation}\label{11}
\big|\varphi_0'\big>=E\otimes\big|0x\big>=\alpha\big|0x_0\big>+\beta\big|1x_1\big>,
\end{equation}
\begin{equation}\label{12}
\big|\varphi_1'\big>=E\otimes\big|1x\big>=m\big|0y_0\big>+n\big|1y_1\big>,
\end{equation}
where $\big|x_i\big>$ and $\big|y_i\big>$ are the pure ancillary states determined by the operation \emph{E} uniquely, and
\begin{equation}\label{13}
|\alpha|^2+|\beta|^2=1, |m|^2+|n|^2=1.
\end{equation}
\indent Then let us calculate the detection probability. Attacked by Eve, the state of composed system becomes
\begin{equation*}
    \big|\psi\big>_{Eve}=I\otimes E\otimes E\otimes E\big[\frac{1}{\sqrt{2}}\big(\big|0x0x0x0x\big>+\big|1x1x1x1x\big>\big)\big]
\end{equation*}
\begin{equation*}
    =\frac{1}{\sqrt{2}}\big[ \big|0\big> \otimes \big(\alpha \big|0x_0\big>+\beta\big|1x_1\big>\big) \otimes \big(\alpha \big|0x_0\big>+\beta\big|1x_1\big>\big) \otimes \big(\alpha \big|0x_0\big>+\beta\big|1x_1\big>\big)
\end{equation*}
\begin{equation*}
    +\big|1\big> \otimes \big(m \big|0y_0\big>+n\big|1y_1\big>\big) \otimes \big(m \big|0y_0\big>+n\big|1y_1\big>\big) \otimes \big(m \big|0y_0\big>+n\big|1y_1\big>\big)\big]
\end{equation*}
\begin{equation*}
    =\frac{1}{\sqrt{2}}\big[\big|0\big> \otimes \big(\alpha^3\big|0x_00x_00x_0\big>
    +\alpha^2\beta\big|0x_00x_01x_1\big>
    +\alpha^2\beta\big|0x_01x_10x_0\big>
    +\alpha\beta^2\big|0x_01x_11x_1\big>
\end{equation*}
\begin{equation*}
    +\alpha^2\beta\big|1x_10x_00x_0\big>
    +\alpha\beta^2\big|1x_10x_01x_1\big>
    +\alpha\beta^2\big|1x_11x_10x_0\big>
    +\beta^3\big|1x_11x_11x_1\big>\big)
\end{equation*}
\begin{equation*}
    +\big|1\big> \otimes \big(m^3\big|0y_00y_00y_0\big>
    +m^2n\big|0y_00y_01y_1\big>
    +m^2n\big|0y_01y_10y_0\big>
    +mn^2\big|0y_01y_11y_1\big>
\end{equation*}
\begin{equation}\label{14}
    +m^2n\big|1y_10y_00y_0\big>
    +mn^2\big|1y_10y_01y_1\big>
    +mn^2\big|1y_11y_10y_0\big>
    +n^3\big|1y_11y_11y_1\big>\big)\big].
\end{equation}
\indent Obviously, when Alice performs measurement on the decoy photons, the probability without eavesdropper is
\begin{equation}\label{15}
    p\big(\big|\psi\big>_{Eve}\big)=\frac{1}{2}\big(|\alpha^3|^2+|n^3|^2\big).
\end{equation}
So the lower bound of the detection probability is
\begin{equation}\label{16}
    d_{lF}=1-p\big(\big|\psi\big>_{Eve}\big)=1-\frac{1}{2}\big(|\alpha^3|^2+|n^3|^2\big).
\end{equation}
Suppose $|\alpha|^2=a, |\beta|^2=b, |m|^2=s, |n|^2=t$, where $a, b,s$ and $t$ are positive real numbers, and $a+b=s+t=1$. Then
\begin{equation}\label{17}
    d_{lF}=1-\frac{1}{2}\big(a^3+t^3\big).
\end{equation}
\indent However, in DPP, authors calculated the efficiency of eavesdropping detection, here don't analyze it again, and the efficiency is
\begin{equation}\label{18}
    d_{lD}=|\beta|^2=|\beta'|^2=1-|\alpha|^2=1-|\alpha'|^2.
\end{equation}
\indent Now, let us analyze how much information Eve can gain maximally when there is no control mode. Similar to that in Ref. [18], first, let us suppose that the quantum state of the photon in the hand of Alice is $\big|0\big>$, Alice takes measurement on the photon in her hand with single-photon detector and the state is $\big|0\big>$. Then the state of the system composed of Bob's photon is
\begin{equation}\label{19}
    \big|\psi'\big>=E\big|0,E\big> \equiv E\big|0\big>\big|E\big>
    =\alpha\big|0\big>\big|\varepsilon_{00}\big>+\beta\big|1\big>\big|\varepsilon_{01}\big>
    \equiv \alpha\big|0,\varepsilon_{00}\big>+\beta\big|1,\varepsilon_{01}\big>,
\end{equation}
and Eve's probe can be described by
\begin{equation}\label{20}
    \rho'=|\alpha|^2\big|0,\varepsilon_{00}\big>\big<0,\varepsilon_{00}\big|
            +|\beta|^2\big|1,\varepsilon_{01}\big>\big<1,\varepsilon_{01}\big|
            +\alpha\beta^*\big|0,\varepsilon_{00}\big>\big<1,\varepsilon_{01}\big|
            +\alpha^*\beta\big|1,\varepsilon_{01}\big>\big<0,\varepsilon_{00}\big|.
\end{equation}
After encoding of the unitary operations $U_0$,$U_1$,$U_2$ and $U_3$ with the probabilities $p_0, p_1,p_2$ and $p_3$, respectively, the state reads
\begin{equation*}
    \rho''=(p_0+p_3)|\alpha|^2\big|0,\varepsilon_{00}\big>\big<0,\varepsilon_{00}\big|
        +(p_0+p_3)|\beta|^2\big|1,\varepsilon_{01}\big>\big<1,\varepsilon_{01}\big|
\end{equation*}
\begin{equation*}
    +(p_0-p_3)\alpha\beta^*\big|0,\varepsilon_{00}\big>\big<1,\varepsilon_{01}\big|
    +(p_0-p_3)\alpha^*\beta|1,\varepsilon_{01}\big>\big<0,\varepsilon_{00}\big|
\end{equation*}
\begin{equation*}
    +(p_1+p_2)|\alpha|^2\big|1,\varepsilon_{00}\big>\big<1,\varepsilon_{00}\big|
    +(p_1+p_2)|\beta|^2\big|0,\varepsilon_{01}\big>\big<0,\varepsilon_{01}\big|
\end{equation*}
\begin{equation}\label{21}
    +(p_1-p_2)\alpha\beta^*\big|1,\varepsilon_{00}\big>\big<0,\varepsilon_{01}\big|
    +(p_1-p_2)\alpha^*\beta\big|0,\varepsilon_{01}\big>\big<1,\varepsilon_{00}\big|,
\end{equation}
which can be rewritten in the orthogonal basis $\{\big|0, \varepsilon_{00}\big>, \big|1,\varepsilon_{01}\big>, \big|1,\varepsilon_{00}\big>, \big|0,\varepsilon_{01}\big>\}$,
\begin{equation}\label{22}
    \rho''=\left(\begin{array}{cccc}
              (p_0+p_3)|\alpha|^2 & (p_0-p_3)\alpha\beta^* & 0 & 0 \\
              (p_0-p_3)\alpha^*\beta & (p_0+p_3)|\beta|^2 & 0 & 0 \\
              0 & 0 & (p_1+p_2)|\alpha|^2 & (p_1-p_2)\alpha\beta^* \\
              0 & 0 & (p_1-p_2)\alpha^*\beta & (p_1+p_2)|\beta|^2
            \end{array}
   \right),
\end{equation}
with
\begin{equation}\label{23}
    p_0+p_1+p_2+p_3=1.
\end{equation}
The information $I_0$ that Eve can get is equal to the Von Neumann entropy
\begin{equation}\label{24}
    I_0=\sum_{i=0}^3 -\lambda_i \log_2 \lambda_i.
\end{equation}
Where $\lambda_i(i=0,1,2,3,)$ are the eigenvalues of $\rho''$, which are
\begin{equation*}
    \lambda_{0,1}=\frac{1}{2}(p_0+p_3) \pm \frac{1}{2}\sqrt{(p_0+p_3)^2-16p_0p_3|\alpha|^2|\beta|^2}
\end{equation*}
\begin{equation}\label{25}
    =\frac{1}{2}(p_0+p_3) \pm \frac{1}{2}\sqrt{(p_0+p_3)^2-16p_0p_3(d-d^2)}
\end{equation}
\begin{equation*}
    \lambda_{2,3}=\frac{1}{2}(p_1+p_2) \pm \frac{1}{2}\sqrt{(p_1+p_2)^2-16p_1p_2|\alpha|^2|\beta|^2}
\end{equation*}
\begin{equation}\label{26}
    =\frac{1}{2}(p_1+p_2) \pm \frac{1}{2}\sqrt{(p_1+p_2)^2-16p_1p_2(d-d^2)}
\end{equation}
In the case of $p_0=p_1=p_2=p_3=0.25$, where Alice encodes exactly 2 bits, expression(25-26) simplify to $\lambda_0=0.5d, \lambda_1=0.5(1-d), \lambda_2=0.5d$ and $\lambda_3=0.5(1-d)$. Interestingly, the maximal information gain is equal to the Shannon entropy of a binary channel
\begin{equation*}
    I_0(d)=-\frac{1}{2}d \log_2(\frac{1}{2}d)-(\frac{1}{2}-\frac{1}{2}d)\log_2(\frac{1}{2}-\frac{1}{2}d)
\end{equation*}
\begin{equation}\label{27}
    -\frac{1}{2}d \log_2(\frac{1}{2}d)-(\frac{1}{2}-\frac{1}{2}d)\log_2(\frac{1}{2}-\frac{1}{2}d).
\end{equation}
\indent Then assume that Bob sends $\big|1\big>$ rather than $\big|0\big>$. The above security analysis can be done in full analogy, resulting in the same crucial relations. The maximal amount of information is equal to the Shannon entropy of a binary channel
\begin{equation*}
    I_1(d)=-\frac{1}{2}d \log_2(\frac{1}{2}d)-(\frac{1}{2}-\frac{1}{2}d)\log_2(\frac{1}{2}-\frac{1}{2}d)
\end{equation*}
\begin{equation}\label{28}
    -\frac{1}{2}d \log_2(\frac{1}{2}d)-(\frac{1}{2}-\frac{1}{2}d)\log_2(\frac{1}{2}-\frac{1}{2}d).
\end{equation}
So the maximal amount of information that Eve can obtain is
\begin{equation}\label{29}
    I=\frac{1}{2}(I_0+I_1)=1-d \log_2d-(1-d) \log_2(1-d).
\end{equation}
After some simple mathematical calculations in FPP, when $a=t$, get
\begin{equation}\label{30}
    d_{lF}=1-a^3,
\end{equation}
and the maximum $I$ is
\begin{equation}\label{31}
    I(d_{lF})=1+H(\sqrt[3]{1-d_{lF}}),
\end{equation}
where
\begin{equation}\label{32}
    H(x)=-x \log_2x-(1-x)\log_2(1-x).
\end{equation}
\indent However, in DPP, the maximum $I$ is
\begin{equation}\label{33}
    I(d_{lD})=1-d_{lD} \log_2d_{lD}-(1-d_{lD})\log_2(1-d_{lD})=1+H(d_{lD}).
\end{equation}
\indent The above analysis shows that function $I(d_{lD})$ and $I(d_{lF})$ have the similar algebraic properties. If Eve wants to gain the full information ($I=2$), the probabilities of eavesdropping detection are $d_{lD}(I=2)=0.5$ in DPP and $d_{lF}(I=2)=0.88$ in FPP.\\
In order to contrast the two functions, Fig.2 is given. As are shown in Fig.2, if Eve wants to gain the full information, she must face a larger detection probability in FPP than DPP. This also indicates that FPP is more secure than DPP.\\
\indent Taking into account the probability \emph{c} of the decoy mode, the effective transmission rate, i.e. the number of message bits per protocol run, is 1-\emph{c}, which is equal to the probability for a message transfer. So, if Eve wants to eavesdrop one message transfer without being detected, the probability for this event is
\begin{equation}\label{34}
    s(c,d)=(1-c)+c(1-d)(1-c)+c^2(1-d)^2(1-c)+...=\frac{1-c}{1-c(1-d)}.
\end{equation}
Then the probability of successful eavesdropping $I=nI(d)$ bits is $s(I,c,d)=s(c,d)^{I/I(d)}$. So
\begin{equation}\label{35}
    s(I,c,d)=\left(\frac{1-c}{1-c(1-d)} \right)^{I/I(d)},
\end{equation}
where
\begin{equation}\label{36}
    I(d)=1+H(\sqrt[3]{1-d}).
\end{equation}
\begin{figure}[H]
  \centering
  \includegraphics[width=7cm]{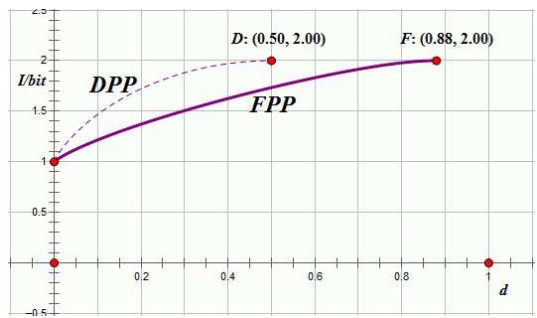}\\
  \caption{The comparison of the two detection results. The dotted line expresses the function $I(d_{lD})$ in DPP and the thick line expresses the function $I(d_{lF})$ in FPP. Obviously, if Eve wants to get the full information, she must encounter the higher detection efficiency in FPP}\label{FIG2}
\end{figure}
\indent Now let us analyze the security of the FPP. In the limit $I\to \infty$ (a message or key of infinite length) get $s \to 0$, so the presented protocol in this paper is \emph{asymptotically secure}. If the security of the quantum channel is ensured, the protocol is completely secure. For example, a choice of the decoy mode is $c=0.5$. In Fig.3, the eavesdropping success probability as a function of the information gain \emph{I} is plotted, for $c=0.5$ and for different detection probabilities \emph{d} which Eve can choose. Note that for $d <0.5$, Eve only gets part of the message right and does not even know which part. So, the FPP protocol is proved secure.
\begin{figure}[H]
  \centering
  \includegraphics[width=7cm]{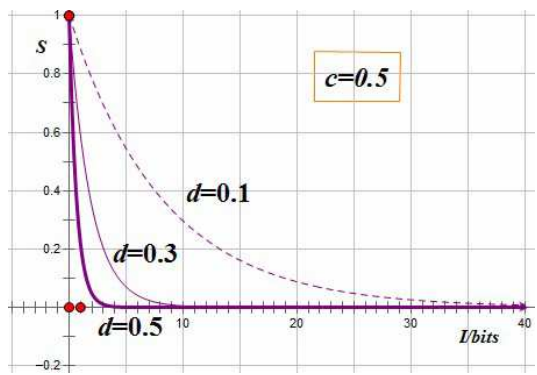}\\
  \caption{Eavesdropping success probability as a function of the maximal eavesdropped information, plotted for different detection probabilities \emph{d}.}\label{FIG3}
\end{figure}
\section{Conclusion and Further Work}
\noindent In summary, an improved eavesdropping detection strategy based on quantum direct communication protocol based on four-particle GHZ state has been introduced, and two eavesdropping detection strategies are compared quantitatively by using the constraint between the information that eavesdropper obtains and the interference introduced. In FPP, the sequence \emph{B} is always in hands of Bob and Eve can only touch the sequence\emph{ A}, and any useful message will not be leaked to the potential eavesdropper. So the security message can be securely transmitted to the receiver. Compared with the DPP, in the FPP protocol, the four-particle GHZ state particles are used to detect eavesdropping which increases the efficiency of detection eavesdropping.\\
\indent In the analysis, if the eavesdropper obtains the full information, she must face a larger detection probability in the FPP than DPP, which shows that the efficiency of eavesdropping detection in FPP is higher than DPP, so it can ensure the quantum direct communication protocol more secure. In order to detect eavesdropping, Bob sends more decoy photons than DPP, while this method reduces the number of measurement. That is, Bob gains the better security at the cost of sending more particles.\\
\indent As we know, the quantum direct communication protocol can also be used as an efficient QKD protocol. In this paper, only the situation that the improved protocol is used as a QKD strategy is considered. So the weaknesses which the quantum direct communication protocol must be faced, such as the noise channel [39-40], the Dos attack [41-42] and so on, may not be considered. In the further work, the other QSDC protocol will be researched.

\end{document}